\documentclass[rnote]{aa}
\bibpunct{(}{)}{;}{a}{}{,} 
\usepackage[varg]{txfonts,epsfig}
\usepackage{natbib}
\usepackage{graphicx}
\usepackage{amsmath}
\begin{document}

\title{Study of the atmospheric conditions at Cerro Armazones using astronomical data\thanks{Based on archival observations collected at the European Organisation for Astronomical Research
in the Southern Hemisphere, Chile and of the Cerro Armazones Observatory facilities of the Ruhr Universit{\"a}t Bochum.}}

\author{Ma\v{s}a Laki\'cevi\'c\inst{1}
\and Stefan Kimeswenger\inst{1,2}
\and Stefan Noll\inst{2}
\and Wolfgang Kausch\inst{3,2}
\and Stefanie Unterguggenberger\inst{2}
\and Florian Kerber\inst{4}}

\institute{Instituto de Astronom{\'i}a, Universidad Cat\'olica del Norte, Av. Angamos 0610, Antofagasta, Chile
\protect{\newline} \email{mlakicevic@ucn.cl, skimeswenger@ucn.cl}
\and Institute for Astro- and Particle Physics, University of Innsbruck, Technikerstr. 25\/8, 6020 Innsbruck, Austria
\and University of Vienna, Department of Astrophysics, T\"urkenschanzstr. 17 (Sternwarte), 1170 Vienna, Austria
\and European Southern Observatory, Karl-Schwarzschild-Str. 2, 85748 Garching bei M{\"u}nchen, Germany}

\date{Received 10 November 2015} 

\abstract {}
   {We studied the precipitable water vapour (PWV) content near Cerro Armazones and discuss the potential use of our technique of modelling the telluric absorbtion lines
for the investigation of other molecular layers. The site is designated for the European Extremely Large Telescope (E-ELT) and
   the nearby planned site for the {\v C}erenkov Telescope Array (CTA).}
   {Spectroscopic data from the Bochum Echelle Spectroscopic Observer (BESO) instrument were investigated by using {line-by-line radiative transfer
   model (LBLRTM)} radiative transfer models for the Earth's atmosphere with the telluric absorption correction tool {\tt molecfit}. All observations from the archive
   {in the period from} December 2008 to the end of 2014 were investigated. {The dataset completely covers the El Ni{\~n}o event registered in the period 2009-2010.} Models of
   the 3D Global Data Assimilation System (GDAS) were used for further comparison. Moreover, for those days with coincidence of data from a similar study
   with VLT/X-shooter and microwave radiometer LHATPRO data at Cerro Paranal, a direct comparison is presented.}
   {{This analysis shows that the site has systematically lower PWV values, even after accounting for the decrease in PWV expected from the higher altitude of the site
   with respect to Cerro Paranal, using the average atmosphere found with radiosondes.} We found that GDAS data are not a suitable method for predicting of local atmospheric conditions -- they usually systematically overestimate the PWV values. Due to the large sample, we were furthermore
   able to characterize the site with respect to symmetry across the sky and variation with the years and within the seasons. This kind of technique of studying
   the atmospheric conditions is shown to be a promising step into a possible monitoring equipment for CTA.}
  {}

\keywords{Atmospheric effects
-- Site testing
-- Instrumentation: miscellaneous
-- Techniques: spectroscopic
-- Methods: observational, numerical
}

\authorrunning{Laki\'cevi\'c et al.}
\titlerunning{Study of the atmospheric conditions at Cerro Armazones using astronomical data}
\maketitle

\section{Introduction}
\sloppy
Site selection and characteristics are often based on long--term campaigns with mainly ground--based facilities \citep{GTC98,TMT1,ELT1, ELT2},
calibrated by a few in situ verifications. These campaigns
measure meteorological parameters like temperature, humidity, wind speed and direction, ambient \hyphenation{pressure}pressure near the surface and seeing
conditions. All these facilities are calibrated on other sites with already existing \hyphenation{enhanced}enhanced astronomical facilities, assuming
similar \hyphenation{ratios}ratios of ground layer to general properties of the atmosphere \citep{ELT3ground}. In the context of site
testing for Giant Magellan Telescope (GMT), Thirty Meter Telescope (TMT) and European Extremely Large Telescope (E-ELT), the determination of atmospheric
precipitable water vapour (PWV) from optical absorption spectroscopy has been established as a standard approach \citep{Querel11}.

Another innovation was the
use of microwave radiometers deployed at several potential sites \citep{otarola10, Kerber10pwv, Kerber12pwv}.
{Multi-wavelength} radiometers, {as e.g. the Low Humidity And Temperature PROfiling microwave radiometer  (LHATPRO\footnote{\tiny Radiometer Physics GmbH, \url{http://www.radiometer-physics.de/}}) used at ESO,} are capable of providing profiles  of the distribution of water vapour in the atmosphere.
The commonly accepted\hyphenation{verification} verification for such {measurements
with balloon-borne radiosondes was shown by \citet{Kerber12pwv}. The results} agree well with {those of } the
radiosondes. Hence, further measurements can be verified with radiometer data directly (at least in desert zones like Cerro
Paranal). The LHATPRO is a radiometer measuring at a frequency of 183 GHz and is hence especially suited
for operations in extremely dry conditions such as the Atacama desert \citep{Rose}. It has been deployed on Paranal 
as a tool to support VLT science operations and has been operational since 2012 \citep{Kerber12pwv}.
Its measurement have been instrumental in documenting episodes of extremely low PWV \citep[well below 0.5 mm,][]{Kerber14mnras}.

\begin{figure*}[ht!]
   \sidecaption
   \includegraphics[width=12cm,bb= 0 0 791 318]{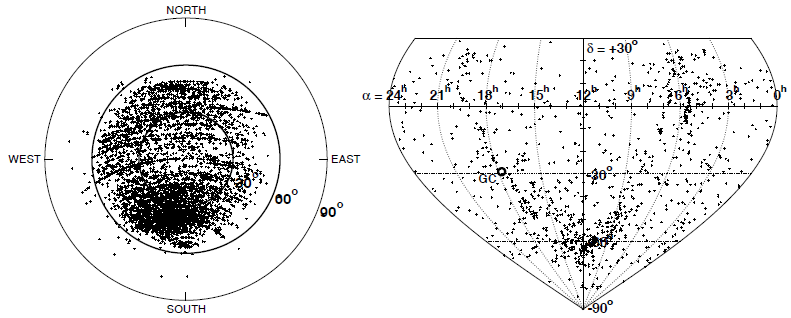}
   \caption{{\it Left panel:} The distribution of the sample sources on the sky. The vast majority of sources was taken
   at zenith distances below 60$\degr$ ($\equiv$ airmass $< 2.0$). {\it Right panel:} The source distribution on the celestial sphere in Eckert-Greifendorff projection. The sources are fairly homogeneously distributed with
   only a slight enhancement towards the Galactic plane. The Galactic center is marked with a circle.}
   \label{FigVibStab}
   \end{figure*}

For the making of local atmospheric models, based on the astronomical data, all the above mentioned facilities are combined. {The global models,} from worldwide weather and climate data calculations (e.g. 3D Global Data Assimilation System
(GDAS)\footnote{\tiny\url{http://www.ready.noaa.gov/gdas1.php}} models) {are used as starting point for the fitting procedures on the astronomical data, which scale the modelled molecular abundance profiles \citep{Smette15}}. This allows the determination of individual absorption of molecules
like ozone, required e.g. for the data reduction of experiments like the {\v C}erenkov Telescope Array (CTA). Furthermore, this allows one to characterize a
site for its homogeneity and asymmetries in the water vapour content caused e.g. by dominating local ground layer air streams, {which is important for observations
with the E-ELT in the infrared spectral bands}. An even more sophisticated solution was presented recently, using local refinement calculations around the
site \citep{Lascaux15,Valparaiso11,Valparaiso15}. {The numerical weather prediction models, in particular those running at high horizontal and vertical
resolution, require extensive computing time, and they are not always readily available to support the operations and astronomical observations program.}

\citet{Noll12} designed a detailed sky model of Cerro Paranal, for ESO.
In a follow--up project, a software {\tt molecfit} was designed for the Earth's lower atmosphere in local thermodynamic equilibrium \citep{Smette15}. {{\tt Molecfit}} is
applied to the astronomical spectroscopic data to determine the atmospheric conditions at the time of the observation. It uses the radiative transfer
code LNFL/LBLRTM\footnote{\tiny\url{http://rtweb.aer.com/lblrtm_frame.html}} \citep{CLO05}. The latter incorporates the spectral line parameter
database {\tt aer\_v\_3.2}, based on HITRAN\footnote{\tiny\url{http://www.cfa.harvard.edu/hitran/}} \citep{Rothman_Hitran}. {Furthermore, {\tt molecfit}
uses a model of the altitude-dependent chemical composition of the atmosphere obtained from a combination of} the ESO MeteoMonitor\footnote{\tiny\url{http://archive.eso.org/asm/ambient-server}}, a standard
atmosphere\footnote{\tiny\url{http://www.atm.ox.ac.uk/RFM/atm/}} and the GDAS model by the National Oceanic and Atmospheric Administration (NOAA;
time-dependent profiles of the temperature, pressure, and humidity). The code calculates the amount of atmospheric molecules by scaling the atmospheric
profiles iteratively with the radiative transfer code LNFL/LBLRTM \citep{CLO05}. By fitting the telluric absorption features of the theoretical spectrum
to the observed one, a final best-fit profile is achieved, which is assumed to be representative to the state of the atmosphere at the time of observation.
This finally allows one to determine the column densities of the incorporated molecular species.

In \cite{Kausch14} we showed the verification of the results against the same \mbox{LHATPRO}, which was used for
a comparison with the balloon experiments in \cite{Kerber12pwv}. {We found that the median difference between the PWV from {\tt molecfit} and from \mbox{LHATPRO} was
 0.12\,mm with $\sigma\,=\,$0.35\,mm}.
The program package is designed for a robustness against inaccurate initial values.
Thus, it is suited for an automatic batch mode capability, as included for pipelines.
Moreover, it is able to run 'on the fly' in monitoring during
operations; e.g. as proposed in \citet{TMT2015} for TMT on Mauna Kea.  \cite{Kimeswenger15} thus proposed a
small telescope with a fiber optic spectrograph for a site of the {\v C}erenkov Telescope Array (CTA) operations.

We present, to our knowledge, for the first time a spatially resolved study for the designated site for the E-ELT which is also
nearby the recently
proposed location for the {\v C}erenkov Telescope Array (CTA). Moreover our study covers six years and includes {El Ni{$\tilde {\rm n}$}o} \cite[a quasi periodic change of the south pacific climate;][]{elnino}. Both were not included in the previous site study for the TMT \citep{otarola10}, covering only about 200 days of radiometer data and containing no information on the spatial distribution. We were able to characterize the water vapor distribution and discuss the potential of
this technique used here for further extensions towards CTA.

\section{The data}

The astronomical observations were made with BESO, a fibre-fed high-resolution
spectrograph at Observatorio Cerro Armazones (OCA), located 22 km from Cerro Paranal in the Atacama desert in
Chile \citep{Steiner06, Steiner08, Hodapp10, Fuhrmann11}. The location is 1.45 km SW (direction 221$\degr$) from the main summit of Cerro Armazones and often called Cerro Murphy, at
an altitude of 2817~m, about 200 m below the \mbox{E-ELT} site. The proposed CTA site is 15.4 km SW (direction 227$\degr$) from our observatory.
The instrument covers wavelengths
from 370 to 840 nm, with a mean instrumental {spectral resolution of about R$\sim$50\,000}. The products of a dedicated data reduction
pipeline \citep{Fuhrmann11} were used
for this study. These fully reduced data sets were provided by the Bochum team. The spectra from 2008 to 2010 were taken at the 1.5 m Hexapod
telescope. Since 2011, spectra were taken at the 0.83 m IRIS telescope. The observations cover typically 16 days on average monthly. The observation days were
selected not only by means of best weather conditions, but mainly by the science cases of the observers such as
deriving orbits and radial velocities of binary stars. These data are without spectrophotometric
calibration.
The data selection was obtained by an automated statistical quality check during the analysis itself, which is described
in the following section in detail.
There are some differences and
inhomogeneities of the data obtained with the two telescopes,
such as existing header keywords and the length of the obtained spectrum.
For a small fraction of less than 100 files a unique reconstruction of observing time, airmass, etc. was not possible, and thus they were rejected.

In total, data from 996 nights out of the period from 2008 to 2014 exist. The observed stars are distributed over the whole sky, with a
slight concentration towards the Galactic plane, covering the sky well until a zenith angle of
$ z \approx 60\degr$. Only a handful of stars exceed this limit. The sample distribution is shown in Fig.~\ref{FigVibStab}.

The precipitable water vapour (PWV) data for the Cerro Paranal sample were taken from \citet{Kausch15}. This data set
contains several months of LHATPRO microwave radiometry data campaign covering various seasons in 2012 and about 3 years of
data taken in the infrared by the VLT instrument X-shooter.

\begin{figure}[ht!]
 \includegraphics[width=87mm,angle=0,bb=0 0 527 411]{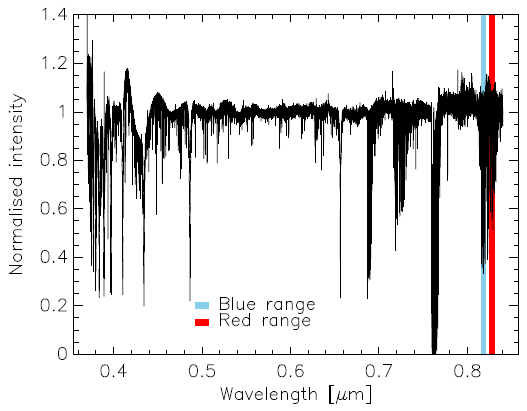}
\caption{An example of whole spectrum taken by BESO; the blue and red fitting region are shown with blue and red colour, respectively.}
 \label{spectrum_example}
 \end{figure}

\section{The methods and the analysis}

\subsection{Analysis of the BESO sample with {\tt molecfit}}

\begin{figure}[ht!]
 \includegraphics[width=88mm,angle=0,bb= 0 0 583 855]{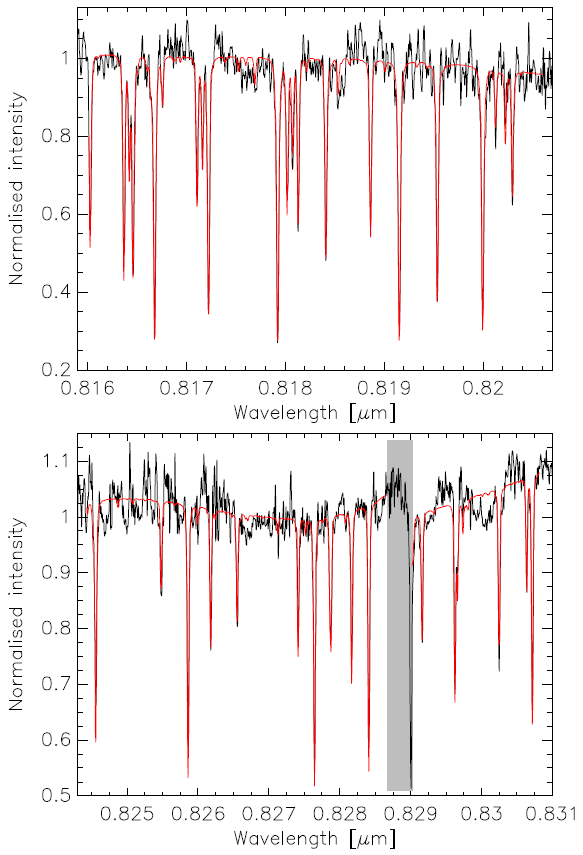}
 \caption{
 {\it Top}: An example of the fitting of the blue range. {\it Bottom}: An example
of the fitting of the red range. The shaded region represents a zone which was excluded from the fit by the {\tt molecfit} code.}
 \label{spectrum}
 \end{figure}

An example of a BESO spectrum is shown in Fig.~\ref{spectrum_example}. The spectra contain telluric absorption bands of
the O$_{3}$, O$_{2}$ and
H$_{2}$O molecules \citep{Smette15}. For each spectra we separately fitted H$_{2}$O in the two narrow ranges, marked in the figure. {{\tt Molecfit}}
calculates the PWV values normalized to zenith.
The fits for the two ranges were obtained independently in order to check for
the quality of the pipeline reduced data.
As these ranges originate from the same populations of the roto-vibrational H$_{2}$O (211$\leftarrow$000) band and thus from the same
volume along the line of sight, they should give
the same results. The so called {\sl blue range} covered the wavelength from 816.01 to 820.60~nm, while the {\sl red range} was defined from 824.43 to
828.67 and from 829.02 to 830.84~nm. In the red range we had to skip the region from 828.67 to 829.02\,nm, due to a strong instrumental artifact which most
likely originates from an overlap of orders in the echelle spectrograph.  The fitting of the sample spectrum for the two ranges is given in Fig.~\ref{spectrum}.
For a detailed description of the methods and how to use {\tt molecfit} see \cite{Smette15}.

\begin{figure}[ht!]
 \includegraphics[width=88mm,angle=0,bb= 0 0 579 404]{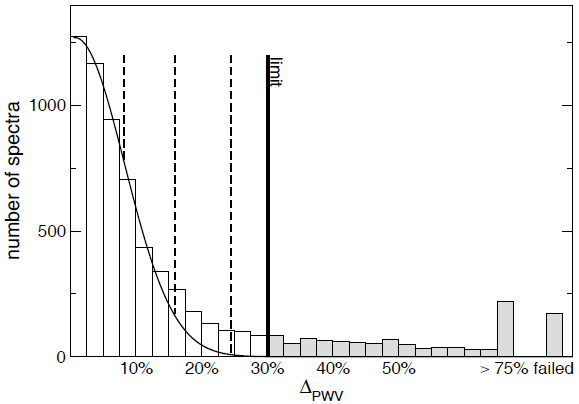}
\caption{The distribution of the differences $\Delta_{\rm PWV}$ and the Gaussian distribution with a {\sl rms} of 7.95\% found
up to the 2$\sigma$ boundary. The dashed lines mark the 1, 2 and 3 $\sigma$ limits respectively.
The shaded part marks the spectra rejected in the sample selection.
Sometimes the fitting of both or one of the ranges completely failed to fit spectral lines. }
 \label{error_hist}
 \end{figure}

During the manual investigation of a small subset of data, to optimize the initial conditions for the
automated fitting routine, some spectra
failed to be fit automatically. This was due to bad background calibration, bad matching of orders of the echelle spectrograph, or some instrument and data reduction pipeline artifacts.
Since for the large data sample a manual selection of the individual fits was not possible, we used a test data set of three months to develop the selection criteria for the automatic rejection of spectra where the fits failed.
In a tradeoff between completeness and reliability we decided towards the conservative approach of higher reliability and thus probably missing
some good data points. After the rejection, the PWV from the blue range was adopted for the final
result (see below), while the red range was only used as a control set. We applied three criteria to select the results:
\begin{enumerate}
\item We selected five narrow windows in both ranges (having in total 108 and 106
spectral pixels, respectively) on top of
the strongest spectral features. They were selected to compare the obtained model and the data{, without having to include a large
amount of pixels where no absorption is applicable}. The latter pixels were only used for {\tt molecfit} to derive the stellar continuum.
The model from fitting the whole range, was compared with
the data only in those windows, building a $\chi^{2}$ value.
We only used those PWV results where the degree of freedom of the fit (5 in our case)
corresponds to a statistical confidence level of $\alpha = 0.05$.
\item We used the {two-sample} Kolmogorov-Smirnov test to check if the observed and fitted spectrum are
identical. We adopted only those results with a confidence level of $\alpha = 0.05$. This test is known
to be very stable for undefined non-normal distribution functions \citep{KS}.
\item
Finally, the differences were calculated as
$$\Delta_{\rm PWV} = \left|{\left({PWV_{\rm red}-PWV_{\rm blue}}\right)/ \left({{\genfrac{}{}{}{2}{1}{2}} \left({PWV_{\rm red}+PWV_{\rm blue}}\right)}\right)\,}\right|$$
and are shown in Fig.~\ref{error_hist}.
The main distribution until the 2$\,\sigma$ level is representing a normal distribution with an {\sl rms} of 7.95\%. Thus, this value is adopted as typical individual error of the measurements.
At larger $\Delta_{\rm PWV}$ systematic errors distort the distribution function.
Only spectra for which the independently derived PWV values from
the two ranges do not differ more than 30\% ($\equiv 3.8\,\sigma$) were selected for the sample.
\end{enumerate}

After applying these three criteria, we obtained PWV estimates based on 5879 spectra, from 918 nights for the final sample.
That corresponds to about 63\% of all archive files. As we still cover 92\% out of all the 996 nights,
no bias towards better PWV by the selection process has been introduced.

\begin{figure}[ht!]
\centerline{\includegraphics[width=70mm,angle=0,bb= 0 0 467 454]{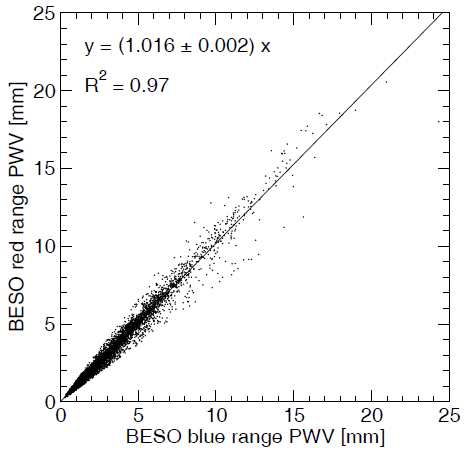}}
\caption{The regression between the results of the red and the blue range fit for
all spectra used in the final analysis. 
} \label{compare3}
\end{figure}

While at normal conditions the values from them are expected to coincide very well, in extreme conditions
(extremely dry and thus weak absorption or extremely wet and thus saturation effects), the values could have larger differences.
In the red fitting range molecular absorptions are weaker. Therefore, that part of {the} spectrum is more difficult to fit in case of dry conditions.
Also, instrumental defects, most likely remnants of fringing, causing periodic wave structures
in the continuum of the star, are stronger in the {red spectral} range. Manual control samples also showed that these can be a source for
some of the larger differences.
After the selection mentioned above based on the control set and the statistical reliability,
we adopted for the further
analysis only the results from the blue range, since the H$_2$O absorption is
stronger there. Furthermore, the range covers a shorter wavelength region in total and
thus is less affected from inter--order errors in the data reduction
of the echelle spectrograph. It is assumed that the fits are more accurate there.
Finally, for the spectra that were not rejected, only an extremely weak systematic effect
between the results of the two ranges was found (Fig.~\ref{compare3}).

\subsection{Homogeneity of the Sky}

\begin{figure}[ht!]
\includegraphics[width=88mm,angle=0,bb= 0 0 558 497]{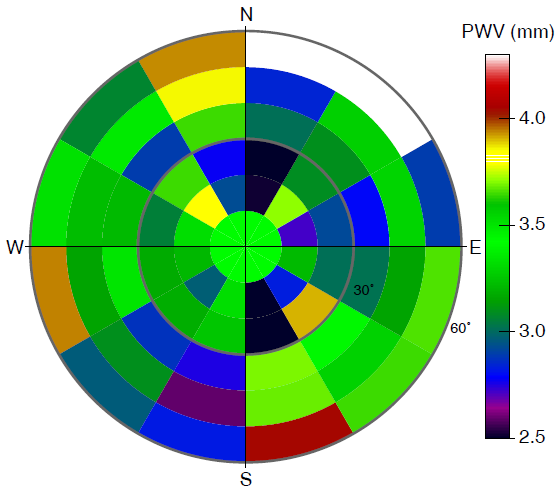}
\caption{The 1.5 $\sigma$ truncated mean PWV for each field on the sky. Note that the bins where {zenith angle is more than 60 degrees} contain
no more than 10 measurements and sometimes only a single one, while there are typically 100 to 200 at all higher sky positions.} \label{mean}
\end{figure}

An important issue for a site is the homogeneity on the sky.
A large number of observations are required for such an
analysis in order to obtain statistically strong results. In our case we could combine data from the full six year period {for separating seasonal and short-term variations from large timescale trends.} 
We tested the spatial homogeneity of the PWV above the site by dividing the sky in
30$^{\circ}$ azimuth and 10$^{\circ}$ zenith distance bins, resulting in a polar coordinate grid of 9$\times$12 fields. As {\tt molecfit} already
recalculates the column density with respect to the zenith solution of the profile, the results are cleaned already from effects due to the airmass. For each bin we calculated a 1.5$\,\sigma$ clipped mean.
{The bins where the zenith distance is less than} 60$^{\circ}$ have an average PWV of 3-4\,mm, and the distribution is fairly homogeneous and independent of the direction (Fig.~\ref{mean}).
Especially no East-West asymmetry can be identified due to the $\sim$35 km distant Pacific coast and/or the air upstreaming the
mountains. Bins with higher zenith distances ($>$60$^{\circ}$) are statistically not reliable, as they contain only 1 to 10
measurements over the
whole period of 6 years and are not used for analysis. The same is true for the two fields where zenith distance is between $50^{\circ}$ and $60^{\circ}$ at an azimuth is
between 0$^{\circ}$ and 60$^{\circ}$, having less than 25 measurements.

These findings are consistent with the results obtained by \citep{Querel14} for Paranal. Using all-sky scans{, obtained every 6 hours over a period of 21
months} (down to $27.5^{\circ}$ elevation), they report that the PWV over Paranal is remarkably uniform with a median variation of 0.32 mm (peak to valley) or 0.07 mm (rms). The homogeneity is a
function of the absolute PWV but the relative variation is fairly constant at 10\% (peak to valley) and 3\% (rms).
Such variations will not have a significant impact on the analysis of astronomical data and they conclude that observations are representative for the whole sky under most conditions.

\subsection{Comparison and Verification with Cerro Paranal}

\begin{figure}[th!]
\includegraphics[width=88mm,angle=0,bb= 0 0 532 408]{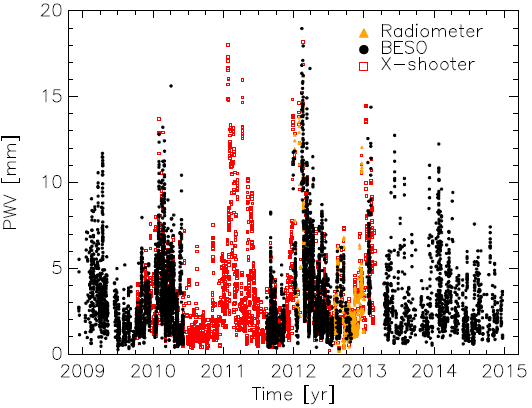}
\caption{The complete data sample from BESO measurements (this work) and the X-shooter and LHATPRO
radiometer data taken from the sample of \citet{Kausch15}. }
\label{sample_compare}
\end{figure}

We compare BESO PWV results with the ones from Cerro Paranal \citep{Kausch15} to verify our results (see Fig.~\ref{sample_compare}). They used X-shooter
data in the same way, by fitting the water vapour bands with {\tt molecfit}. While the typical seasonal
variations follow nicely the same trends at both sites, there is no clear signature of the {moderate-to-strong} {El Ni{$\tilde {\rm n}$}o}
\citep{elnino} during the Chilean summer 2009/2010.
Interestingly enough, \citet{EWASS15} reported an unusually large number of extremely dry ($\le 0.50$\,gr/m$^3$ H$_2$O) and very dry ($\le 0.65$\,gr/m$^3$ H$_2$O) events for the years 2008, 2009 and 2010 by measuring the absolute humidity at 30-m height above Paranal. These events were 2--3 times more frequent than in other years. Thus, it seems that the 2009/2010 {El Ni{$\tilde {\rm n}$}o} event did not cause more humid
conditions at Cerro Paranal.

\begin{figure}[th!]
\includegraphics[width=88mm,angle=0,bb= 0 0 575 330]{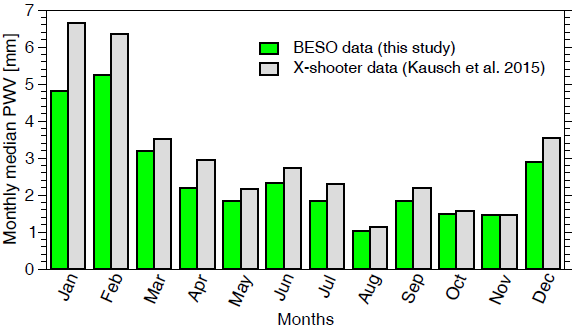}
\caption{
The monthly median values of the PWV at Cerro Paranal and Cerro Murphy, {using only those measurements where the observing times do not
differ more than 1.5 hours}. The subsample sizes are 228, 434, 458, 330, 380, 96, 36, 232, 524, 212, 76, 148, from January
to December, respectively.}
\label{evol}
\end{figure}

To get widely bias--free and comparable data sets, only those spectra of the X-shooter and BESO samples were selected which were taken nearly simultaneously (within 1.5 hours).
This results in 3154 pairs of data. The resulting median distribution as function of the months
for both sites is shown in Fig.~\ref{evol}{, where the uncertainty varies between 4\% and 16\%}. The difference of the two sites is especially obvious
during the more humid summer. For nights with coincident observations the median PWV for Cerro Murphy (BESO) is 2.43 mm, while it is 2.96 mm for Cerro Paranal
(X-shooter). This {difference of 22\%} is significantly
larger than expected by the difference of the altitude (2817 and 2635~m for Cerro Murphy and Cerro Paranal, respectively), {using the scale-height of 1800~m found by balloon-born radiosondes from Antofagasta for the altitudes above 2\,km by \citet{Otarola2011}. As they showed, the scale height in the Atacama desert can be lower under dry conditions. However, a random sampled spot check of the {\tt molecfit} results from our sample with 20 probes around the median PWV revealed that it was well in agreement at these average conditions}. Using {\tt molecfit}, including ground--based meteorological data \citep{Smette15},
with data from Cerro Paranal leads to a theoretical difference of only about 8\% due to the altitude. {The PWV as function of site altitude given by \citet{otarola10} leads to 8.2\%. This was obtained by comparing the 5 tested TMT candidate sites, covering an altitude range from 2290\,m to 4480\,m.}

\begin{figure}[ht!]
\includegraphics[width=88mm,angle=0,bb= 0 0 570 398]{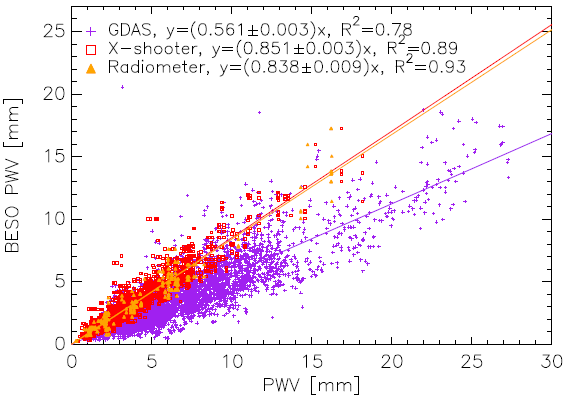}
\caption{Linear fits of X-shooter, radiometer, and GDAS PWV at the abscissa respectively vs. BESO PWV at the ordinate.
}
\label{compall}
\end{figure}

To supplement, we compare the BESO PWV with the values derived from GDAS as well as with those observed within 1.5 hours
with X-shooter and with the verification data set of the LHATPRO radiometer \citep{Kausch14, Kausch15}.
In Fig.~\ref{compall} we analyzed the linear correlation ($y=Ax$) of the BESO data with those from Cerro Paranal. The resulting coefficients and R$^{2}$ values
are given in Table~\ref{table:1}. The BESO and X-shooter data show nearly identical slope (0.85)
as BESO and the verification data set from LHATPRO radiometer (0.84); with
$R^2$ of 0.89 and 0.93, respectively.
Thus, one can conclude that the LHATPRO radiometer verification as well can be used for BESO.

The $R^2$ of BESO with GDAS data is 0.78 (Fig.~\ref{compall}). As we can see, using GDAS data, which works fine for the prediction far from the coast, on
the other side of the Andes mountains, at
the Auger observatory \citep{GDAS_AUGER}, does not work in this zone of the Atacama desert. GDAS predictions systematically overestimate the PWV by 52\%. Similar results were found for Cerro Paranal, but using already locally improved GDAS profiles merged with ground station meteorological data \citep{Kausch15}.

\begin{table}[ht!]
\caption{\label{prva} Comparison of BESO PWV with the ones from Cerro Paranal and GDAS data for Cerro Murphy. Linear fitting $y\,=\,A\,x$ was
performed between pairs of PWV data.}
\label{table:1}
\begin{tabular}{l c c c}
\hline\hline
Data & $A$ &  & R$^2$ \\
\hline
X-shooter & 0.851$\,\,\pm$0.003 &  &0.89 \\
Radiometer & 0.838$\,\,\pm$0.009&  &0.93 \\
GDAS & 0.561$\,\,\pm$0.003 &  & 0.78 \\
\hline
\end{tabular}
\end{table}

\section{Conclusions and Outlook}

\subsection{PWV over Cerro Murphy}
As already shown for Cerro Paranal in \citet{Kausch15}, spectral feature fitting by {\tt molecfit} can  successfully be used to investigate the PWV content in the Earth's atmosphere. The resulting
PWV from Cerro Murphy, southwest of Cerro Armazones, are systematically lower than those from Cerro Paranal. The effect is stronger than expected from the altitude difference
of 182~m{, assuming the water vapor vertical distribution
at the median conditions}. The PWV is
homogeneous and does not show any major sky direction asymmetries or trends within the six years of the study, although it covers the El Ni{\~n}o event 2009/10.
Thus, it is possible to parameterize the sky only by means of zenith distance, without including
azimuth dependencies. We show that a very small telescope facility is able to provide an atmospheric
monitoring capability with results fully comparable to those from the large telescopes. Moreover, it is clear that the interpolated data from GDAS are too inaccurate.
Recently, two independent local refinement calculations of these coarse grid weather models were presented by
\citet{Lascaux15} and \citet{Valparaiso15}. Certainly, a comparison study of these methods with our results would be a further step.
For this
or other studies we could provide our resulting atmosphere profiles electronically.

\subsection{Outlook to atmosphere studies for CTA}
Up to now, data reduction uses only simple integrated profiles through the whole atmosphere as function of the airmass, or
if very high accuracy is required only uses nearby calibrators (e.g. differential photometry or iso-airmass telluric standards).
The latter is not always possible or requires a
large calibration plan.
The first one suffers from missing information on the azimuth dependence, and is not sufficient for describing light propagation of sources located in the atmosphere as it is the case for
$\gamma$-ray astronomy if using {\v C}erenkov radiation.

If we use individual species behaving differently due to distributions other than the classical
exponential scale height, real height profiles are especially important.
The distribution of water vapour is strongly concentrated to the lower 10 km
and thus do not follow these simple dependencies as normally used for wide band filter extinction curves.

Molecules in some distinct layers have to be treated specifically. 
An important example for this is ozone. Nearly all ozone absorption occurs in the stratosphere.
It has to be removed from the
extinction curve derived by photometry \citep{photometer,FRAM} for the CTA data reduction.
%
For measuring the variation of the tropospheric layers, our methods of ground-based spectroscopy presented here will be not sensitive enough.
With the standard profiles of our sky model \citep{Noll12} we calculated the average fraction of the
tropospheric ozone from the ground at 2\,km to 10\,km.
As it only contributes about 6\% to the total ozone absorption
throughout the whole atmosphere, the error
introduced by assuming a constant low fraction of tropospheric ozone on the total calculation is small.
The strong high atmosphere contribution has to be
derived separately from other sources of extinction and subtracted from the overall absorption.

The Chappuis bands at $500 < \lambda < 700$~nm are absorbing and scattering
significantly up to 4\% at zenith.
That is already essential, as the flux calibration specification for CTA is 5-7\%.
However, they only show very shallow and highly blended absorption structures.
Our spectra do not cover the blue to the ultraviolet light sufficiently to also
model the ozone variation in the Huggins bands ($\lambda\,<\,400$~nm), which are stronger and
show more characteristic features. Thus, they would be more appropriate for measuring the total ozone column.
Therefore, a proposed CTA installation should extend
to the ultraviolet until 320~nm to fit the profile from the Huggins bands.
The contribution of the Chappuis bands  can be calculated by our radiative transfer codes thereafter \citep[][Fig. 2]{Kimeswenger15}.


\begin{acknowledgements} We would like to thank the referee for his comments and the very detailed suggestions.
Furthermore we thank Angel Ot{\'a}rola for providing us his full presentation of the EWASS15 conference.
We thank Christian Westhues and Thomas Dembsky from Ruhr Universit{\"a}t Bochum for their
help in accessing the BESO archive. M.L. was funded by the
Comit{\'e} Mixto ESO-Gobierno de Chile in the project {\sl Investigaci{\'o}n Atmosf{\'e}rica con
Instrumentaci{\'o}n Astron{\'o}mica}. {\tt Molecfit} was designed in the framework
of the Austrian ESO In-Kind project funded by BM:wf under contracts
BMWF-10.490/0009-II/10/2009 and BMWF-10.490/0008-II/3/2011. S.N and S.U. are funded by the
Austrian Science Fund (FWF): P26130 and W.K. is supported by the project IS538003
(Hochschulraumstrukturmittel) provided by the Austrian Ministry for Research, Investigation and Economy (BM:wfw).
\end{acknowledgements}

\bibliography{BESOMolecfit_arxiv}

\end{document}